\documentclass[pra,twocolumn,showpacs,amssymb]{revtex4}
\usepackage{graphicx}
\usepackage{amsmath}

\begin{document}

\title{Berezinskii-Kosterlitz-Thouless transition 
\\ in the time-reversal-symmetric Hofstadter-Hubbard model}

\author{M. Iskin}
\affiliation{Department of Physics, Ko\c{c} University, Rumelifeneri Yolu, 
34450 Sar\i yer, Istanbul, Turkey}

\date{\today}

\begin{abstract}

Assuming that two-component Fermi gases with opposite artificial magnetic 
fields on a square optical lattice are well-described by the so-called 
time-reversal-symmetric Hofstadter-Hubbard model, we explore the thermal
superfluid properties along with the critical Berezinskii-Kosterlitz-Thouless
(BKT) transition temperature in this model over a wide-range of its parameters. 
In particular, since our self-consistent BCS-BKT approach takes the 
multi-band butterfly spectrum explicitly into account, it unveils how 
dramatically the inter-band contribution to the phase stiffness dominates 
the intra-band one with an increasing interaction strength for any 
given magnetic flux. 

\end{abstract}


\maketitle

\section{Introduction}
\label{sec:intro}

The phase stiffness, also known as the helicity modulus, measures 
the response of a system in an ordered phase to a twist of the order 
parameter~\cite{fisher73}, and it is directly linked to the superfluid 
(SF) density of the superconducting systems~\cite{scalapino92, scalapino93}. 
In its most familiar form, the conventional expression for the elements of 
the phase stiffness tensor can be written as~\cite{denteneer93}
\begin{align}
D_{\mu\nu}^{oneband} = 
\frac{1}{\mathcal{V}_d} \sum_{\mathbf{k}} 
&\left\lbrace 
\frac{\partial^2 \xi_{\mathbf{k}}}{\partial k_\mu \partial k_\nu}
\left[1 - \frac{\xi_{\mathbf{k}}} {E_{\mathbf{k}}} 
\tanh \left(\frac{E_{\mathbf{k}}}{2k_BT} \right) \right]
\right.
\nonumber \\ 
-
\frac{1}{2k_BT} 
&
\left.
\left(\frac{\partial \xi_{\mathbf{k}}}{\partial k_\mu} 
\frac{\partial \xi_{\mathbf{k}}}{\partial k_\nu} \right)
\textrm{sech}^2 \left( \frac{E_{\mathbf{k}}}{2k_BT} \right)
\right\rbrace,
\label{eqn:conv}
\end{align}
where $\mathcal{V}_d$ is the volume element and $\mathbf{k}$ is the 
wave vector in $d$ spatial dimensions,
$k_\nu$ with $\nu \equiv \{x,y,..\}$ is the projection of $\mathbf{k}$,
$\xi_\mathbf{k} = \varepsilon_\mathbf{k} - \mu$
is the single-particle dispersion relation shifted by the chemical potential,
$E_\mathbf{k} = \sqrt{\xi_\mathbf{k}^2 + |\Delta_\mathbf{k}|^2}$ 
is the quasi-particle dispersion relation with the order parameter $\Delta_\mathbf{k}$,
$k_B$ is the Boltzmann constant, and $T$ is the temperature. Here, 
$\varepsilon_\mathbf{k}$ is assumed to be quite general, and not limited 
with the usual quadratic dependence on $k_\nu$.
In particular, this tensor plays a special role in two dimensions for which it 
appears explicitly in the universal BKT relation determining the critical SF 
transition temperature $T_{BKT}$~\cite{scalapino92, scalapino93,denteneer93}.
This is a topological phase transition characterized by the binding (unbinding) 
of two vortices with opposite circulations, i.e., the so-called vortex-antivortex
pairs, below (above) $T_{BKT}$ with algebraically (exponentially) 
decaying spatial correlations~\cite{b, kt, nk}. 
For instance, one of the immediate manifestations of Eq.~(\ref{eqn:conv}) 
is that it rules out the possibility of superfluidity in 
systems with a nearly flat $\mathbf{k}$-space dispersion, 
i.e., the SF density/current is identically zero since the particles are strictly 
immobile in a flat-band with $\xi_{\mathbf{k}} \approx \xi_{0}$ 
for all $\mathbf{k}$.

Motivated by the experimental advances with cold Fermi gases, the 
calculation of $D_{\mu\nu}$ have recently been extended to a class of 
multi-band Hamiltonians that are characterized by a single mean-field 
order parameter $\Delta$ with a uniform spread in real space, 
and that exhibit time-reversal ($\mathcal{T}$) 
symmetry~\cite{torma15, torma16, torma17a, torma17b}. It has been found that,
in addition to the intra-band contribution to $D_{\mu\nu}$ that has 
exactly the same form as the one given in Eq.~(\ref{eqn:conv}) for each 
single-particle band, the inter-band contribution may also be necessary
for a proper description of the multi-band systems. For instance, in marked 
contrast with the single flat-band systems, it turns out that superfluidity 
may prevail in a flat-band in the presence of other bands as a result of 
the inter-band tunnelings~\cite{torma15, torma16}. See also the related 
discussion on two-band superconductivity in graphene for a resolution 
of the `superconductivity without supercurrent' controversy in the vicinity 
of its Dirac points~\cite{kopnin08, kopnin10}.

In view of the recent realization of the Hofstadter-Hubbard model with 
$\mathcal{T}$ symmetry~\cite{bloch13trs, ketterle13trs}, and the forthcoming
experiments, here we study $T_{BKT}$ in this model and address the 
interplay between the intra-band and inter-band contributions to the 
phase stiffness in the presence of a multi-band butterfly spectrum. 
Despite our naive expectations, we find that the maximum 
$T_{BKT} \approx 0.253t/k_B$ is attained for the no-flux limit at $\mu = 0$ 
when the interaction strength is around $U \sim 3.75t$. 
Here, $t$ is the hopping strength.
In addition, one of the highlights of this paper is that increasing the 
interaction strength always shifts the relative importance of the intra-band 
and inter-band contributions in an overwhelming favor of the latter, and 
that the proper description of the Cooper molecules requires an indiscriminate 
account of both contributions in the strong-coupling limit. 

The remainder of this paper is organized as follows. After a short overview 
of the Hofstadter model with $\mathcal{T}$ symmetry in Sec.~\ref{sec:butterfly},
first we introduce the self-consistent BCS-BKT formalism in Sec.~\ref{sec:bkt},
together with the multi-band generalization of the phase stiffness detailed 
in Sec.~\ref{sec:stiffness}. Then we discuss the analytically-tractable strong-coupling
or molecular limit in Sec.~\ref{sec:molecular} as a warm of for our numerical 
results presented in Sec.~\ref{sec:numerics}. We end the paper with a brief 
summary of our conclusions in Sec.~\ref{sec:conc}.

\section{Theoretical Framework}
\label{sec:theory}

Assuming that the tight-binding approximation is a viable description 
of the kinematics of a two-component Fermi gas on an optical lattice, 
we start with the single-particle Hamiltonian
$
H_0 = - \sum_{ij} c^\dag_{i} t_{ij} c_{j},
$
where $c^{\dagger}_{i}$ ($c_{i}$) creates (annihilates) a spinless fermion 
at site $i$ so that $t_{ij} = t_{ji}^*$ is the element of the hopping matrix from 
site $j$ to $i$. This model also offers a convenient way to incorporate the 
effects of additional gauge fields, e.g., an external magnetic field
$
\mathbf{B}(\mathbf{r}) = \nabla \times \mathbf{A}(\mathbf{r})
$ 
may be taken into account via the minimal coupling, i.e.,
$
t_{ij} \to t_{ij} e^{\mathrm{i} 2\pi \phi_{ij}},
$
leading to an additional phase factor 
$
\phi_{ij} = (1/\phi_0) \int_{\mathbf{r}_j}^{\mathbf{r}_i} 
\mathbf{A}(\mathbf{r}) \cdot d\mathbf{r}
$
in the hopping matrix. Here, $\phi_0$ is the flux quantum
and $\mathbf{A}(\mathbf{r})$ is the magnetic vector potential. In a 
broader context, $\mathbf{A}(\mathbf{r})$ could be any gauge field,
including the artificial ones created in atomic systems.

In this paper, we are interested in a square lattice lying in the entire 
$(x,y)$ plane, that is under the influence of a spatially-uniform 
magnetic field $B (\mathbf{r}) = B$ pointing along the perpendicular 
$z$ axis. Such a setting can be represented by 
$
\mathbf{A}(\mathbf{r}) = (0, B x, 0)
$ 
in the Landau gauge without losing generality. Thus, for a given flux 
quanta per unit cell $\alpha = Ba^2/\phi_0$ with $a$ the lattice spacing, 
the particle gains an Aharonov-Bohm phase $e^{\mathrm{i} 2\pi \alpha}$ 
after traversing a loop around the unit cell. Next we consider the original 
Hofstadter model~\cite{hofstadter76} and allow the particle to hop back 
and forth between only the nearest-neighbor sites, which by itself gives 
rise to one of the most fascinating single-particle energy spectra in nature.

\subsection{Hofstadter Butterfly}
\label{sec:butterfly}

When the flux $\alpha = p/q$ corresponds precisely to a ratio of 
two relatively prime numbers $p$ and $q$, one can simplify the single-particle 
problem considerably by switching to the reciprocal ($\mathbf{k}$) space 
representation and making use of the new translational 
symmetry~\cite{hofstadter76}. That is, since the $B$ field enlarges the unit 
cell by a factor of $q$ in the $x$ direction, the first magnetic 
Brillouin zone (MBZ) is reduced to
$
-\pi/(qa) \le k_x < \pi/(qa)
$
and
$
-\pi/a \le k_y < \pi/a.
$
This not only splits the tight-binding $s$-band of the flux-free system, i.e., 
$
\varepsilon_\mathbf{k} = -2t \cos(k_x a) - 2t \cos(k_y a),
$
into $q$ sub-bands for a given $\alpha = p/q$, but the competition between
the magnetic length scale, i.e., the cyclotron radius, and the periodicity of the 
lattice potential also produces a very complicated energy $\varepsilon$ 
versus $p/q$ landscape with an underlying fractal pattern. 
As the overall landscape bears a resemblance to the shape of a butterfly, 
the spectrum is usually referred to as the Hofstadter butterfly in the 
literature~\cite{hofstadter76}.

For a given $\alpha = p/q$, the multi-band spectrum can be obtained by 
solving the Schr\"odinger equation
$
\mathbb {H}_{0 \mathbf{k}} |n \mathbf{k} \rangle 
= \varepsilon_{n \mathbf{k}} |n \mathbf{k} \rangle
$
in $\mathbf{k}$ space, where $n = 0,\ldots,q-1$ labels the sub-bands starting 
from the lowest-energy branch. This leads to
$
\sum_{j=0}^{q-1} H_{0\mathbf{k}}^{ij} g_{n \mathbf{k}}^{j} 
= \varepsilon_{n \mathbf{k}} g_{n \mathbf{k}}^{i},
$
where $g_{n \mathbf{k}}^i$ is the $i = 0,1,\cdots,q-1$th component 
of the $n=0,1,\cdots,q-1$th eigenvector of the single-particle problem 
with energy $\varepsilon_{n \mathbf{k}}$. Here, the Fourier-expansion 
coefficient $c_{i \mathbf{k}}$ of the site operator can be written in terms 
of the band operators $d_{n \mathbf{k}}$ as
$
c_{i \mathbf{k}} = \sum_{n=0}^{q-1} g_{n \mathbf{k}}^i d_{n \mathbf{k}}.
$
Thus, for a given $\mathbf{k}$ in the MBZ, the spectrum is determined 
by solving the following equation,
\begin{align}
\left(
\begin{array}{cccccc}
  B_{\mathbf{k}}^0 & C_\mathbf{k} & 0 & . & 0 & C_\mathbf{k}^* \\
  C_\mathbf{k}^* & B_{\mathbf{k}}^1 & C_\mathbf{k} & 0 & . & 0 \\
  0 & \ddots & \ddots & \ddots & 0 & . \\
  . & 0 & C_\mathbf{k}^* & B_{\mathbf{k}}^j & C_\mathbf{k} & 0  \\
  0 & . & 0 & \ddots & \ddots & \ddots \\
  C_\mathbf{k}& 0 & . & 0 & C_\mathbf{k}^* & B_{\mathbf{k}}^{q-1} \\
\end{array}
\right)
\left(
\begin{array}{c}
g_{n \mathbf{k}}^{0} \\
. \\
g_{n \mathbf{k}}^{j-1} \\
g_{n \mathbf{k}}^{j}  \\
g_{n \mathbf{k}}^{j+1}  \\
. \\
g_{n \mathbf{k}}^{q-1} \\
\end{array}
\right)
= 0,
\label{eqn:hofsmatrix}
\end{align}
where 
$
B_{\mathbf{k}}^j = -2t\cos(k_ya + 2\pi \alpha j) 
- \varepsilon_{n \mathbf{k}}
$
with $j = 0,1,\cdots,q-1$ and $C_\mathbf{k} = -te^{\mathrm{i} k_x a}$. 
We note that while the ortho-normalization condition
$
\langle n \mathbf{k} | m \mathbf{k} \rangle = \delta_{nm}
$
leads to
$
\sum_{j=0}^{q-1} g_{n \mathbf{k}}^{j*} g_{m \mathbf{k}}^{j} = \delta_{nm},
$
the completeness relation
$
\sum _{n=0}^{q-1} | n \mathbf{k} \rangle \langle n \mathbf{k} | = \mathbb{I}
$
leads to
$
\sum_{n=0}^{q-1} g_{n \mathbf{k}}^{i*} g_{n \mathbf{k}}^{j} = \delta_{ij}
$
for any given $\mathbf{k}$ state, both of which are used throughout the paper 
in simplifying the self-consistency equations.

The butterfly spectrum exhibits a number of symmetries. First of all, it preserves 
the inversion symmetry in $\mathbf{k}$ space, i.e.,
$
\varepsilon_{n \mathbf{k}} = \varepsilon_{n, -\mathbf{k}},
$
as a direct manifestation of the gauge invariance in a uniform flux.
In addition, it is not only symmetric around $\varepsilon = 0$ for a given flux, i.e.,
$
\varepsilon_{n \mathbf{k}} (\alpha) = - \varepsilon_{q-1-n, -\mathbf{k}} (\alpha),
$
due to the particle-hole symmetry of $H_0$ on a bipartite lattice, 
but it is also mirror-symmetric around $\alpha = 1/2$ for a given 
$|n \mathbf{k} \rangle$ state, i.e.,
$
\varepsilon_{n \mathbf{k}} (\alpha) = \varepsilon_{n\mathbf{k}} (1-\alpha)
$ 
for $ 0 \le \alpha \le 1$.
The latter relation suggests that the flux-free $\alpha = 0$ system is exactly 
equivalent to the $\alpha = 1$ case, and hence, $\alpha = 1/2$ corresponds 
to the maximally attainable flux~\cite{hofstadter76}. 
Furthermore, when $q$ is an even denominator, combination of the inversion 
and particle-hole symmetries implies the condition
$
\varepsilon_{q/2-1,\mathbf{k}} = -\varepsilon_{q/2,\mathbf{k}},
$
from which we infer that the centrally-symmetric bands $n = q/2-1$ and 
$n = q/2$ have degenerate $\mathbf{k}$ states with $\varepsilon = 0$. 
It turns out that these central bands contain $q$ Dirac cones with $q$ 
zero-energy touchings in the first MBZ, and therefore, are not separated 
by a bulk energy gap. For example, setting $p/q = 1/2$, we obtain
$
\varepsilon_{n \mathbf{k}} = (-1)^{n+1} 2t\sqrt{\cos^2(k_x a) + \cos^2(k_y a)}
$
for the $n = 0$ and $n = 1$ bands, where
$
g_{n \mathbf{k}}^0 / g_{n \mathbf{k}}^1 
=  -\cos(k_y a) / [\cos(k_y a) + (-1)^{n+1} \sqrt{\cos^2(k_x a) + \cos^2(k_y a)}
$
together with 
$
|g_{n \mathbf{k}}^0|^2 + |g_{n \mathbf{k}}^1|^2 = 1.
$
The locations of the Dirac points are
$
k_x = -\pi/(2a)
$
and
$
k_y = \pm \pi/(2a).
$
Since $\alpha = 1/2$ case is a non-trivial yet an analytically tractable one, 
we often use it as one of the ultimate benchmarks for the accuracy of 
our numerical calculations.

Our primary interest in this paper is the superfluidity of a spin-$1/2$ Fermi gas 
on a square lattice that is experiencing an equal but opposite magnetic fields
for its spin components~\cite{troyer14, umucalilar17, iskin17}, 
i.e., $\alpha_\uparrow = - \alpha_\downarrow = \alpha$. 
This restores the $\mathcal{T}$ symmetry into the system in such a way 
that the solutions of the Hofstadter model for a $\downarrow$ particle can 
be written in terms of the $\uparrow$ ones (given above) as follows:
$
\varepsilon_{\downarrow n \mathbf{k}} = \varepsilon_{\uparrow n, -\mathbf{k}}
$
and
$
g_{\downarrow n \mathbf{k}}^{j} = g_{\uparrow n,-\mathbf{k}}^{j^*}.
$
More importantly, the self-consistent mean-field theory of such a 
time-reversal-symmetric Hofstadter-Hubbard model~\cite{umucalilar17, iskin17} 
turns out to be dramatically simpler to implement than that of the usual 
Hofstadter-Hubbard model~\cite{oktel10, umucalilar16}, as we discuss next.

\subsection{Self-consistent BCS-BKT Theory}
\label{sec:bkt}

Having an equal number of $\uparrow$ and $\downarrow$ particles that are
interacting with on-site and attractive interactions in mind, we have recently 
shown that the mean-field Hamiltonian can be simply written 
as~\cite{umucalilar17, iskin17}
\begin{eqnarray}
H = \sum_{\sigma n \mathbf{k}} 
\xi_{n \mathbf{k}} d^\dag_{\sigma n \mathbf{k}} d_{\sigma n \mathbf{k}}
- \Delta \sum_{n \mathbf{k}} \left( d_{\uparrow n \mathbf{k}}^\dag 
d_{\downarrow n, -\mathbf{k}}^\dag +\text{H.c.} \right ),
\label{eqn:ham}
\end{eqnarray}
in $\mathbf{k}$ space. Here, $H$ is given up to a constant $M \Delta^2/U$ 
term, where $M = \mathcal{A}/a^2$ is the number of lattice sites with 
$\mathcal{A}$ the area of the system, 
$
\xi_{n \mathbf{k}}=\varepsilon_{n \mathbf{k}} - \mu
$ 
is the butterfly spectrum $\varepsilon_{n \mathbf{k}}$ shifted by the chemical 
potential $\mu$, and
$
\Delta = (U/M) \sum_{n\mathbf{k}} 
\langle d_{\downarrow n,-\mathbf{k}} d_{\uparrow n \mathbf{k}} \rangle
$ 
is the order parameter characterizing a spatially-uniform SF phase. 
In addition, $U \ge 0$ is the strength of the inter-particle interactions, 
$\langle \ldots \rangle$ denotes the thermal average, 
$\text{H.c.}$ is the Hermitian conjugate, and $\Delta$ is assumed 
to be real without losing generality.
In contrary to the usual Hofstadter-Hubbard model where the competing 
vortex-lattice like SF phases involve both intra- and inter-band Cooper 
pairings with nontrivial sets of finite center of mass momenta $\mathbf{K}$, 
and hence, require $q \times q$ order parameters~\cite{oktel10, umucalilar16}, 
here the energetically more favorable mean-field solution boils down to 
the superfluidity of intra-band Cooper pairs with $\mathbf{K} = \mathbf{0}$ 
only~\cite{umucalilar17, iskin17}. This is simply 
because, as the $\mathcal{T}$ symmetry guarantees the existence of a 
$\downarrow$ partner in state $|n,-\mathbf{k} \rangle$ for every 
$\uparrow$ fermion in state $|n \mathbf{k}\rangle$, the spatially-uniform 
SF solution allows all particles to take advantage of the attractive 
potential by making $\uparrow \downarrow$ Cooper pairs with 
$\mathbf{K} = \mathbf{0}$. We also emphasize that the 
disappearance of all of the inter-band pairing terms from the mean-field 
Hamiltonian is a direct consequence of the uniform SF phase with 
$\mathcal{T}$ symmetry.

Given the quadratic Hamiltonian, minimization of the corresponding 
thermodynamic potential with respect to $\Delta$, together with the 
number equation
$
N = \sum_{\sigma n \mathbf{k}}
\langle d_{\sigma n \mathbf{k}}^\dag d_{\sigma n \mathbf{k}} \rangle
$
that is controlled by $\mu$, leads to a closed set of self-consistency 
equations that are analytically tractable. For instance, a compact 
way to express these mean-field 
equations is~\cite{umucalilar17, iskin17, umucalilar17note}
\begin{align}
\label{eqn:op}
1 &= \frac{U}{2M} \sum_{n\mathbf{k}} \frac{\mathcal{X}_{n\mathbf{k}}}{E_{n\mathbf{k}}}, \\
F &= 1 - \frac{1}{M} \sum_{n\mathbf{k}} 
\frac{\mathcal{X}_{n\mathbf{k}}}{E_{n\mathbf{k}}} \xi_{n\mathbf{k}},
\label{eqn:filling}
\end{align}
where
$
\mathcal{X}_{n\mathbf{k}} = \tanh [E_{n\mathbf{k}}/(2k_B T)]
$
is a thermal factor with $k_B$ the Boltzmann constant and $T$ the temperature,
$
E_{n\mathbf{k}} = \sqrt{\xi_{n\mathbf{k}}^2 + \Delta^2}
$
is the energy spectrum of the quasiparticles arising from band $n$, and the 
particle filling $0 \le F = N/M\le 2$ corresponds to the total number of particles 
per site. Thus, we use Eqs.~(\ref{eqn:op}) and~(\ref{eqn:filling}) to determine 
$\Delta$ and $\mu$ for any given set of $U$, $F$, $T$ and $\alpha$ parameters. 

Since neither the amplitude nor the phase fluctuations of the SF order 
parameter are included in the mean-field theory, while the self-consistent
solutions of Eqs.~(\ref{eqn:op}) and~(\ref{eqn:filling}) is a reliable 
description of the ground state at $T = 0$ for all $U$ values, the theory 
works reasonably well at finite temperatures $T \lesssim T_{BCS} \ll t/k_B$ 
as long as $U \lesssim t$ is weak~\cite{gapbcs}. 
Here, $T_{BCS}$ is the critical BCS 
transition temperature that is determined by setting $\Delta \to 0$.
However, as the role played by the temporal phase fluctuations increases 
dramatically with stronger $U \gtrsim t$ values, the mean-field theory becomes 
gradually insufficient, failing eventually at capturing the finite temperature 
correlations of the SF phase in the strong-coupling or molecular limit even 
though $T \ll T_{BCS}$~\cite{nsr85, randeria92}.
In the $U \gg t$ limit, we note that the mean-field $T_{BCS}$ is directly proportional 
to the binding energy $U$ of the two-body bound state in vacuum, 
and therefore, it characterizes the pair formation temperature of the Cooper 
molecules. Thus, away from the weak-coupling limit, $T_{BCS}$ has obviously 
nothing to do with the critical SF transition temperature of the system, for which 
the phase coherence is known to be established at a much lower temperature.

Taking only the phase fluctuations into account in our two dimensional 
model characterized by a single SF order parameter, the critical SF transition 
temperature is determined by the universal BKT 
relation~\cite{b, kt, nk, denteneer93, bktnote}
\begin{align}
k_B T_{BKT} = \frac{\pi}{8} D_0(T_{BKT}),
\label{eqn:bkt}
\end{align}
where $D_0$ is the isotropic measure of the $2 \times 2$ phase 
stiffness tensor,
i.e., 
$
D_{\mu\nu} = D_0 \delta_{\mu\nu}
$ 
with
$
(\mu, \nu) \equiv \lbrace x, y\rbrace.
$ 
Similar to the usual Hubbard model with a single SF order parameter~\cite{denteneer93}, 
and thanks to the time-reversal symmetry of the current model, the elements 
$D_{\mu\nu}$ are identified by making an analogy with the effective phase-only 
XY Hamiltonian~\cite{torma15, torma17a}, where
$
H_{XY} = (1/8) \int dx \int dy \sum_{\mu\nu} 
\partial_{\mu} \theta_\mathbf{r} D_{\mu\nu} \partial_{\nu} \theta_\mathbf{r}
$
under the assumption that 
$
\Delta_\mathbf{r} = \Delta e^{i\theta_\mathbf{r}}.
$
Setting $\theta_\mathbf{r} = \mathbf{K} \cdot \mathbf{r}$ for a 
spatially-uniform condensate density with $\hbar \mathbf{K}$ the pair 
momentum, we note that
$
H_{XY} = D_0 \mathcal{A} K^2 / 4 = m_0 \mathcal{A} \rho_s v^2/2,
$
where $\mathbf{v} = \hbar \mathbf{K}/(2m_0)$ is the velocity of the SF pairs
with $m_0$ the mass of the particles, and $\rho_s = m_0 D_0/\hbar^2$ 
is the density of the SF particles. Thus, the phase stiffness of a SF is 
essentially equivalent to its SF density. Here, the factor $m_0 \rho_s$ 
is often called the SF mass density of the system. 

We note that since $T_{BCS}$ is determined by the BCS condition 
$\Delta \to 0$ and a finite $T_{BKT}$ requires a finite $\Delta$ by definition,
Eq.~(\ref{eqn:bkt}) already puts $T_{BCS}$ as the upper bound on 
$T_{BKT}$ for any $U \ne 0$. It turns out that while 
$T_{BKT} \sim t^2/(k_B U) \ll T_{BCS}$ in the $U/t \gg 1$ 
limit~\cite{nsr85, randeria92}, 
$T_{BKT} \to T_{BCS}$ in the opposite $U/t \lesssim 1$ limit where 
the rate $D_0/t \to 0$ is the same as $\Delta/t \to 0$ only when $U/t \to 0$.
In fact, we find in Sec.~\ref{sec:numerics} that the maximum 
$T_{BKT} \approx 0.253t/k_B$ is attained for $q = 1$ at $\mu = 0$
when $U \sim 3.75t$.

Next we justify that the self-consistent solutions of 
Eqs.~(\ref{eqn:op})-(\ref{eqn:bkt}) for $\Delta$, $\mu$ and $T_{BKT}$, 
along with the proper multi-band generalization~\cite{torma17a} of 
$D_{\mu \nu}$ given below, is a reliable description of the SF transition 
temperature $T_{BKT}$ for any given set of $U$, $F$ and $\alpha$ 
parameters. We note in passing that the self-consistent BCS-BKT approach 
amounts to be the simultaneous solutions of BCS mean-field equations 
and the universal BKT relation, i.e., the phase fluctuations are 
taken only into account by the latter via the analogy with the underlying 
XY model. While this simple description is known to be quite accurate 
for the weak-coupling BCS and strong coupling molecular limits, it  
provides a qualitative but reliable picture of the crossover regime.

\subsection{Phase Stiffness}
\label{sec:stiffness}

As an alternative to the expression given in Ref.~\cite{torma17a}, a compact 
way to write the elements of the phase stiffness tensor is
\begin{align}
D_{\mu\nu} &= \frac{\Delta^2}{\mathcal{A}} \sum_{nm\mathbf{k}} 
\left(
\frac{\mathcal{X}_{n\mathbf{k}}}{E_{n\mathbf{k}}}
- \frac{\mathcal{X}_{n\mathbf{k}}-\mathcal{X}_{m\mathbf{k}}}{E_{n\mathbf{k}}-E_{m\mathbf{k}}}
\right) \nonumber \\
&\hspace{2cm} \times \frac{2 Q_{\mu\nu}^{nm\mathbf{k}}} 
{E_{m\mathbf{k}}(E_{n\mathbf{k}}+E_{m\mathbf{k}})},
\label{eqn:stiffness}
\end{align}
where the independent $n$ and $m$ summations run over all bands, 
$
\mathcal{X}_{n\mathbf{k}} = \tanh [E_{n\mathbf{k}}/(2k_B T)]
$
is the thermal factor, and the coefficient
$
Q_{\mu\nu}^{nm\mathbf{k}} = \textrm{Re} 
[
\langle n \mathbf{k}| 
\partial \mathbb{H}_{0 \mathbf{k}} / \partial k_\mu
|m \mathbf{k} \rangle
\langle m \mathbf{k}| 
\partial \mathbb{H}_{0 \mathbf{k}} / \partial k_\nu
| n \mathbf{k} \rangle
]
$
is directly related to the details of the band geometry of the single-particle 
problem~\cite{kopninnote}.
For instance, since $D_{\mu \nu}$ is isotropic in space for spatially-uniform 
SFs, we have 
$
Q_{\mu \nu}^{nm\mathbf{k}} 
= Q_{0}^{nm\mathbf{k}} \delta_{\mu \nu}
$
with the particular coefficient
$
Q_{0}^{nm\mathbf{k}} = 4t^2 a^2 \left| \sum_{j = 0}^{q-1} \sin(k_y a+ 2\pi \alpha j) 
g_{n \mathbf{k}}^{j*} g_{m \mathbf{k}}^j \right|^2
$
obtained for our model Hamiltonian given in Eq.~(\ref{eqn:hofsmatrix}).
Furthermore, by denoting
$
D_{\mu\nu} = D_{\mu\nu}^{intra} + D_{\mu\nu}^{inter},
$ 
we distinguish the intra-band contribution of the phase stiffness from the 
inter-band contribution, that is based, respectively, on whether $n = m$ or 
not in Eq.~(\ref{eqn:stiffness}). Such an association proves to be illuminating 
in some of our analysis given below.

First, let us show that the intra-band contribution of Eq.~(\ref{eqn:stiffness}) 
corresponds precisely to the conventional expression given in 
Eq.~(\ref{eqn:conv}). Setting $n \to m$ for the intra-band contribution, 
the second term in the parenthesis implies a derivative such that 
$
d\tanh(a x)/dx = a \textrm{sech}^2(a x),
$ 
and the coefficient 
$
Q_{\mu\nu}^{nn\mathbf{k}} = 
(\partial \varepsilon_{n\mathbf{k}}/\partial k_\mu) 
(\partial \varepsilon_{n\mathbf{k}}/\partial k_\nu) 
$
depends only on the spectrum. After plugging them into Eq.~(\ref{eqn:stiffness}), 
we rearrange the intra-band contribution into two pieces as
$
D_{\mu\nu}^{intra} = (1/\mathcal{A}) \sum_{n\mathbf{k}} 
(\partial \xi_{n\mathbf{k}}/\partial k_\mu) 
\partial (\xi_{n\mathbf{k}} \mathcal{X}_{n\mathbf{k}}/E_{n\mathbf{k}})/\partial k_\nu
-[1/(2\mathcal{A}k_B T)] \sum_{n\mathbf{k}} \textrm{sech}^2[E_{n\mathbf{k}}/(2k_BT)]
(\partial \xi_{n\mathbf{k}}/\partial k_\mu) 
(\partial \xi_{n\mathbf{k}}/\partial k_\nu).
$
Since the latter piece already appears in the conventional expression, next we
recast the first piece into two summations as
$
(1/\mathcal{A}) \sum_{n\mathbf{k}} 
\partial[(\partial \xi_{n\mathbf{k}}/\partial k_\mu) \xi_{n\mathbf{k}} 
\mathcal{X}_{n\mathbf{k}}/E_{n\mathbf{k}}]/\partial k_\nu
-(1/\mathcal{A}) \sum_{n\mathbf{k}} [\partial^2 \xi_{n\mathbf{k}}/(\partial k_\mu \partial k_\nu)]
\xi_{n\mathbf{k}} \mathcal{X}_{n\mathbf{k}}/E_{n\mathbf{k}}.
$
Note here that while the first summation integrates to zero as the derivatives 
$\partial \xi_{n\mathbf{k}}/\partial k_\mu$
vanish at the MBZ boundaries, the second summation is equivalent to
$
(1/\mathcal{A}) \sum_{n\mathbf{k}} (1 - \xi_{n\mathbf{k}} \mathcal{X}_{n\mathbf{k}}/E_{n\mathbf{k}})
[\partial^2 \xi_{n\mathbf{k}}/(\partial k_\mu \partial k_\nu)],
$
as the additional summation
$
(1/\mathcal{A}) \sum_{n\mathbf{k}} \partial^2 \xi_{n\mathbf{k}}/(\partial k_\mu \partial k_\nu) = 0
$
integrates to zero for the same reason given just above. Thus, the intra-band 
contribution of Eq.~(\ref{eqn:stiffness}) is precisely the conventional 
expression in disguise.

For a more explicit demonstration, we substitute 
$
\varepsilon_{\mathbf{k}} = \hbar^2 k^2/(2m_0)
$
for the dispersion relation in Eq.~(\ref{eqn:stiffness}), and obtain the SF stiffness
of a single-band continuum system at $T = 0$ as,
$
D_{\mu \nu} = [\hbar^2 \Delta^2/(m_0^2 \mathcal{A})] 
\sum_{\mathbf{k}} k_\mu k_\nu / E_{\mathbf{k}}^3,
$
leading to
$
D_{0} = (\mu + \sqrt{\mu^2 + \Delta^2}) /(2\pi)
$ 
for any $\Delta \ne 0$.
After plugging the mean-field solutions
$
\Delta = \sqrt{2\varepsilon_b \varepsilon_F}
$
and 
$
\mu = \varepsilon_F - \varepsilon_b/2
$ 
into this expression, where $\varepsilon_F = \hbar^2 k_F^2/(2m_0)$ 
is the Fermi energy and $\varepsilon_b \ge 0$ is the two-body binding 
energy in vacuum, i.e., 
$
1 = (U/M) \sum_{\mathbf{k}} 
1/(2\varepsilon_{\mathbf{k}} + \varepsilon_b),
$
we obtain
$
D_{0}= \hbar^2 \rho_F / m_0
$
for any $\Delta \ne 0$. Here, $\rho_F  = N/\mathcal{A} = k_F^2/(2\pi)$ is the total density 
of particles with $k_F$ the Fermi wave vector. Alternatively, this result follows 
immediately from the $T = 0$ limits of the conventional expression
$
D_{0} = [\hbar^2/ (m_0 \mathcal{A})] \sum_{\mathbf{k}} (1-\xi_{\mathbf{k}}/ E_{\mathbf{k}}),
$
given in Eq.~(\ref{eqn:conv}), together with the number equation
$
N = \sum_{\sigma \mathbf{k}} [1/2-\xi_{\mathbf{k}}/ (2E_{\mathbf{k}})].
$
Thus, $D_0$ suggests that the entire continuum Fermi gas becomes a SF 
for any $\Delta \ne 0$, i.e., $\rho_s = \rho_F$ at $T = 0$ as soon as $U > 0$.
By making an analogy with this continuum result, we identify the SF 
density of particles for the lattice model in Sec.~\ref{sec:molecular}.

Second, in the case of two-band SFs, e.g., when $q = 2$ in our model
as discussed in Sec.~\ref{sec:butterfly}, it can be explicitly shown 
that~\cite{torma17a} the integrand of the inter-band contribution is linked 
to the total quantum metric of the bands
$
\sum_{n=\pm} f_{\mu \nu}^{n \mathbf{k}}
$
with 
$
f_{\mu \nu}^{+,\mathbf{k}} = f_{\mu \nu}^{-,\mathbf{k}},
$
where
$
f_{\mu \nu}^{n \mathbf{k}} = \textrm{Re}
[\langle \partial_{k_\mu} n \mathbf{k} | 
(\mathbb{I} - | n \mathbf{k} \rangle \langle n \mathbf{k} |) 
| \partial_{k_\nu} n \mathbf{k} \rangle]
$
or equivalently
$
f_{\mu \nu}^{n \mathbf{k}} = \sum_{m \lbrace \ne n \rbrace} 
Q_{\mu \nu}^{nm\mathbf{k}} / 
(\varepsilon_{n \mathbf{k}} - \varepsilon_{m \mathbf{k}})^2
$
is the quantum metric of band $n$ in general.
Note that, as the eigenfunctions $g_{n \mathbf{k}}^{j}$ are only determined 
up to a random phase factor for a given $|n \mathbf{k} \rangle$
in a computer program, their partial derivatives contain indefinite factors, 
making the former expression unsuitable for numerical computation. 
This ambiguity is nicely resolved by transforming the derivatives to 
the Hamiltonian matrix in the latter expression.

Third, the general expression given in Eq.~(\ref{eqn:stiffness}) acquires a 
much simpler form at sufficiently low temperatures when $k_BT \ll \Delta$.
Assuming this is the case, we set $\mathcal{X}_{n\mathbf{k}} \to 1$ for 
every $|n \mathbf{k}\rangle$ state, and obtain
\begin{align}
D_{\mu\nu} &= \frac{\Delta^2}{\mathcal{A}} \sum_{nm\mathbf{k}} 
\frac{2 Q_{\mu\nu}^{nm\mathbf{k}}} 
{E_{n\mathbf{k}} E_{m\mathbf{k}}(E_{n\mathbf{k}}+E_{m\mathbf{k}})}.
\label{eqn:stiffness0}
\end{align}
To be exact, this expression is precisely the $T = 0$ limit of 
Eq.~(\ref{eqn:stiffness}). However, it is also valid for all 
$T \le T_{BKT}$ in the molecular limit, since $k_B T_{BKT} \sim t^2/U$ 
when $U/t \gg 1$, and hence, $k_BT_{BKT} \ll t \ll \Delta$ is well-founded.

\subsection{Molecular Limit}
\label{sec:molecular}

In the $\Delta \gg t$ or equivalently $U \gg t$ limit of tightly-bound Cooper
molecules~\cite{nsr85, randeria92}, Eqs.~(\ref{eqn:op}) and~(\ref{eqn:filling}) 
give
$
\Delta = (U/2) \sqrt{F(2-F)}
$
and
$
\mu = -(U/2) (1-F),
$
so that $\sqrt{\mu^2 + \Delta^2} = U/2$ is independent of $T$.
Therefore, both of these mean-field parameters are not only proportional 
to the binding energy $U$ of the two-body bound state in 
vacuum, but are also independent of $\alpha$ 
as the $\mathcal{T}$ symmetry ensures that the 
center of mass of the Cooper pairs are neutral against the flux. 
In other words, the only mechanism that allows 
a Cooper molecule to hop from one site to another is via the virtual breaking 
of its $\uparrow$ and $\downarrow$ constituents~\cite{nsr85, randeria92}. 
Since the cost for breaking the bound state is $U$, the molecule effectively
hops from site $j$ to site $i$ with
$
t_{m ij} = 2 t_{\uparrow ij} t_{\downarrow ij}/U.
$
Thus, for our nearest-neighbor lattice model, we identify 
$
t_m = 2t^2/U
$ 
as the hopping amplitude of the molecules, and
$
\alpha_m = \alpha_\uparrow + \alpha_\downarrow = 0
$
as their flux.

In addition, this intuition further suggests that the SF density, 
and hence the SF phase stiffness, must be independent of $\alpha$ 
in the molecular limit. We prove this physical expectation by first 
approximating Eq.~(\ref{eqn:stiffness0}) as,
$
D_{\mu\nu} = \lbrace \Delta^2/[\mathcal{A} (\mu^2+\Delta^2)^{3/2}]\rbrace 
\sum_{nm\mathbf{k}} Q_{\mu\nu}^{nm\mathbf{k}},
$
and then noting that the summations over $n$ and $m$ is equivalent to
$
\sum_{nm} Q_{\mu\nu}^{nm\mathbf{k}}
= \textrm{Tr}
[
(\partial \mathbb{H}_{0 \mathbf{k}} / \partial k_\mu)
(\partial \mathbb{H}_{0 \mathbf{k}} / \partial k_\nu)
],
$
for any given $\mathbf{k}$. In particular to our Hamiltonian given in 
Eq.~(\ref{eqn:hofsmatrix}), we immediately get
$
4t^2  \sum_{\mathbf{k} \in \textrm{MBZ}} \sum_{j = 0}^{q-1} 
\sin^2(k_ya+2\pi\alpha j) 
$
for the $\mu \nu \equiv yy$ element, and this summation is exactly equivalent to
that of the flux-free system, i.e.,
$
4t^2 \sum_{\mathbf{k} \in \textrm{BZ}} \sin^2(k_y a) = M/2,
$
since the interval $-\pi \le k_y a < \pi$ remains unchanged in both Brillouin zones.
Thus, we conclude that 
$
D_0 = 4F(2-F) t^2/U
$
is independent of both $\alpha$ and $T \le T_{BKT}$ in the molecular limit.
It is worth highlighting that, given the indiscriminate account of both the 
intra-band and inter-band contributions in recovering the desired $D_0$ 
of the Cooper molecules, our proof offers an indirect yet an impartial 
support of the recent results. In fact, our numerical calculations presented 
in Sec.~\ref{sec:numerics} reveal that $D_0^{inter}$ eventually 
dominates over $D_0^{intra}$ with increasing $U/t$ for any $q \ge 2$. 
While this domination is substantial even for the simplest two-band and 
three-band ($q = 2$ and $3$) cases, it is already quite dramatic for 
$q > 3$ as the butterfly bands get more flattened.

In addition, to make an analogy with the molecular limit of a continuum 
Fermi gas that is discussed at length in Sec.~\ref{sec:stiffness}, we first
recall that the expression 
$
D_0 = \hbar^2 \rho_s /m_0
$
is derived for all $\Delta \ne 0$ including the molecular limit. 
Second, we rewrite $D_0$ in terms of the SF density of the continuum 
molecules $\rho_{sm} = \rho_{s}/2$ and their mass $m_{0m} = 2m_0$ as
$D_0 = 4\hbar^2 \rho_{sm}/m_{0m}$.
Then, by plugging the effective mass $m_{0m} = \hbar^2/(2t_m a^2)$ of the 
lattice molecules into this continuum expression, we identify
$
F_{sm} = a^2 \rho_{sm} = U D_0/(16t^2) = (F/2)(1-F/2)
$
as the filling of SF molecules. It is pleasing to confirm that $F_{sm}$ is 
independent of $\alpha$, which need not be the case for the filling 
of SF particles $F_{s} = a^2 \rho_s = D_0/(2t)$ in the weak-coupling limit.

On the other hand, by adapting the definition of the number of condensed 
particles for our model~\cite{leggett},
\begin{align}
F_{c} = \frac{\Delta^2}{2M} \sum_{n\mathbf{k}} 
\frac{\mathcal{X}_{n\mathbf{k}}^2}{E_{n\mathbf{k}}^2},
\end{align}
and taking the molecular limit, we obtain 
$
F_{cm} = \Delta^2/[4(\mu^2+\Delta^2)] = (F/2)(1-F/2)
$
as the filling of condensed molecules, which is also independent of $\alpha$. 
Thus, we conclude that all of the SF molecules are condensed with a fraction 
of $2F_{sm}/F = 2F_{cm}/F = 1-F/2$. In perfect agreement with the 
continuum model where we find that the entire Fermi gas is condensed 
and become SF in the dilute ($F \to 0$) limit, half of the Fermi gas is not 
condensed at half-filling ($F \to 1$). This difference between the dilute 
continuum and finite-filling lattice has to do with the fact that Cooper 
molecules are intrinsically hardcore by their composite nature, which is
strictly dictated by the Pauli exclusion principle in the $U/t \to \infty$ limit. 
For this reason, whether a site is almost empty or singly occupied by 
one of the Cooper molecules gives rise to a notable outcome in 
lattice models.

\begin{figure*}[htbp]
\includegraphics[scale=0.8]{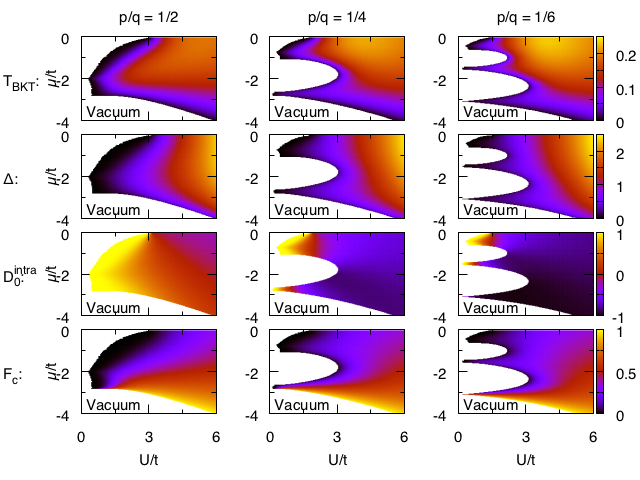}
\label{fig:246}
\caption{(color online)
The critical SF transition temperature $k_B T_{BKT}/t$  is shown in the first row
together with the corresponding SF order parameter $\Delta/t$ in the second row,
relative weight of the intra-band and inter-band contributions 
to the phase stiffness $(D_0^{intra} - D_0^{inter}) / D_0$ in the third row,
and condensate fraction $F_c/F$ in the last row.
}
\end{figure*}
\section{Numerical Results}
\label{sec:numerics}

To illustrate the numerical accuracy of our analysis given in the previous
Sec.~\ref{sec:theory}, next we present the self-consistent solutions of
Eqs.~(\ref{eqn:op})-(\ref{eqn:bkt}) for two sets of $\alpha = 1/q$: the
even $q \in \{2,4,6\}$ set is shown in Fig.~1 and the odd 
$q \in \{3,5,7\}$ is shown in Fig.~2. Here, we primarily 
focus on the evolution of $T_{BKT}$ together with the corresponding 
$\Delta$, $D_0^{intra}$, and $F_c$ in the $\mu$ versus $U$ plane.
The trivial $q=1$ case is included as App.~\ref{sec:app} for the sake
of completeness. Thanks to the particle-hole symmetry of the model 
Hamiltonian, we restrict numerics to $\mu \le 0$ or equivalently $F \le 1$, 
as the solutions are mirror-symmetric around $\mu = 0$ or the half-filling $F = 1$.

First of all, since $T_{BKT}/t \to T_c/t \to 0$ as $\Delta/t \to 0$ or $U \to U_c$, 
where the value of the critical interaction threshold $U_c$ for SF 
pairing depends strongly on the energy density $\mathcal{D}(\varepsilon)$ 
of single-particle states, e.g., $U_c/t > 0$ when $\mu$ lies within the 
butterfly gaps or at $\mu = 0$ when $q$ is even, the top two rows in 
Figs.~1 and~2 recover the overall structure of the ground-state 
($T \to 0$) phase diagrams~\cite{umucalilar17}. 
We recall that the multi-band butterfly spectrum gives rise to a number 
of insulating lobes that are reminiscent of the well-known Mott-insulator 
transitions of the Bose-Hubbard model. This is because while $\Delta$, 
and therefore, $T_{BKT}$ grows exponentially $e^{-1/[U \mathcal{D}(\mu)]}$ 
slow with $U \ne 0$ and $\mathcal{D}(\varepsilon)$ wherever $\mu$ lies 
within any of the butterfly bands, it grows linearly $U-U_c$ fast from 
the semi-metal when $\mu = 0$ and $q$ is even, and with a square 
root $\sqrt{U-U_c}$ from the insulators in general~\cite{umucalilar17, iskin17}.

Even though $\Delta$ and $T_{BKT}$ must, in theory, vanish strictly 
at $U = 0$ wherever $\mu$ lies within any of the butterfly bands, 
this appears not to be the case in any of Figs.~1 
and~2, e.g., $U \to 0$ regions appear white instead of 
black. This is due to a lack of our numerical resolution as the non-linear
solver fails to converge once the relative accuracy of two consecutive 
$\Delta$ iterations reduces below the order of $10^{-5}$. We checked 
that using a $10^{-6}$ resolution does not improve the phase diagrams, 
i.e., the minor corrections are indistinguishable to the eye. 
On the other hand, this shortage makes the general structure 
of $\mathcal{D}(\varepsilon)$ visible on the periphery of the white regions. 
In contrast, the insulating lobes are determined quite accurately, 
since $\Delta$ and $T_{BKT}$ vanish very rapidly as $U \to U_c \ne 0$.

The top rows in Figs.~1 and~2 show that the maximum critical temperatures 
are always attained at $\mu = 0$ for some intermediate $U \sim 3.75t-5.5t$, 
and are all of the order of $k_BT_c^{max} \sim 0.19t-0.25 t$ for any 
given $\alpha$. In particular, we approximately determine the following 
$(k_BT_c^{max}/t, U/t)$ values in our numerics:
$(0.1917, 5.45)$ for $q = 2$,
$(0.2027, 4.70)$ for $q = 3$,
$(0.2181, 4.25)$ for $q = 4$,
$(0.2260, 4.10)$ for $q = 5$,
$(0.2321, 4.00)$ for $q = 6$, and
$(0.2363, 3.95)$ for $q = 7$.
Thus, increasing $q$ from $2$ not only enhances $\max k_B T_c^{max}/t$
quite monotonously, but it also occurs at a lower $U/t$. In comparison, 
we find $(0.2528, 3.75)$ for $q  = \infty$ or equivalently $q = 1$ 
corresponding to the usual no-flux model presented in App.~\ref{sec:app}. 
Assuming that the monotonic trend continues for larger $q$, we suspect 
that the result of $q = \infty$ case is an ultimate upper bound for
$T_c^{max}$ in the entire parameter range of the model Hamiltonian
given in Eq.~(\ref{eqn:ham}). 
In the molecular limit when $U \gg t$, we verify that $k_B T_{BKT}/t$ 
decreases as $\pi F(2-F) t/(2U)$ in all figures, which is in perfect 
agreement with the analysis given above in Sec.~\ref{sec:molecular}.

\begin{figure*}[htbp]
\includegraphics[scale=0.8]{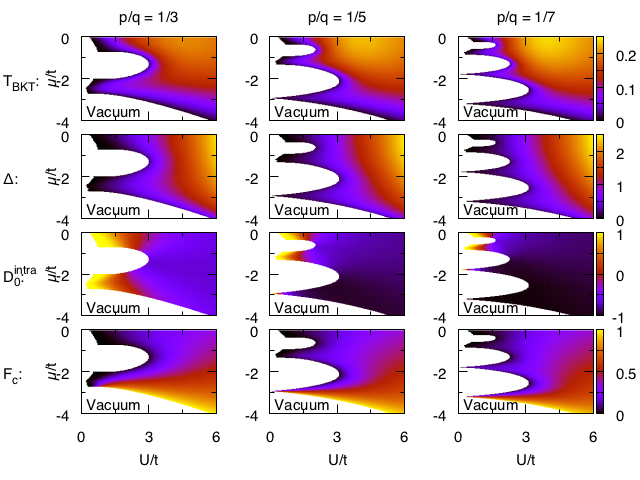}
\label{fig:357}
\caption{(color online)
The critical SF transition temperature $k_B T_{BKT}/t$  is shown in the first row
together with the corresponding SF order parameter $\Delta/t$ in the second row,
relative weight of the intra-band and inter-band contributions 
to the phase stiffness $(D_0^{intra} - D_0^{inter}) / D_0$ in the third row,
and condensate fraction $F_c/F$ in the last row.
}
\end{figure*}

In addition, we present the relative $D_0^{intra} - D_0^{inter}$ weights 
of the intra-band and inter-band contributions to $D_0$ in the third rows 
of Figs.~1 and~2. Together with the $T_{BKT}$ figures shown in the top 
rows which are directly proportional to the sum $D_0^{intra} + D_0^{inter}$,
these results reveal that $D_0^{inter}$ eventually dominates over 
$D_0^{intra}$ with increasing $U/t$ for any $q \ge 2$. 
While this domination is substantial even for the simplest two-band and 
three-band ($q = 2$ and $3$) cases, it becomes sheer dramatic for $q > 3$ 
once the butterfly bands get more flattened. Thus, our numerical
results unveil and highlight the relative importance of $D_0^{inter}$ 
contribution without a doubt.

Lastly, the condensate fractions $F_c/F$ are shown in the bottom rows 
of Figs.~1 and~2. These results show that 
$F_c/F \to 0$ is directly controlled by $\Delta$ in the weak-coupling 
limit when $\Delta/t \to 0$. On the other hand, $F_c/F$ saturates to 
$1 - F/2$ in the molecular limit, which is again in perfect agreement 
with the ground-state analysis given above in Sec.~\ref{sec:molecular}. 
This is because $k_B T_{BKT} /t \to 0$ in both $U \to U_c$ and $U \gg t$ limits.
Having achieved the primary objectives of this paper, next we are ready to 
end it with a brief summary of our conclusions.

\section{Conclusions}
\label{sec:conc}

In summary, by studying the thermal SF properties along with the critical 
SF transition temperature in the Hofstadter-Hubbard model with
$\mathcal{T}$ symmetry, here we analyzed the competition between the 
intra-band and inter-band contributions to the phase stiffness in the 
presence of a multi-band butterfly spectrum. For instance, 
one of the highlights of this paper is that increasing the interaction strength 
always shifts the relative importance of the two in favor of the inter-band
contribution. In marked contrast with the two-band and three-band cases 
for which the shift takes place gradually, our numerical results showed an 
extremely striking shift for the higher-band ones. Last but not least, we also
showed analytically that the proper description of the Cooper molecules 
requires an indiscriminate account of both contributions in the 
strong-coupling limit. 

Given our convincing evidence that the inter-band effects are absolutely
non-negligible in a typical multi-band butterfly spectrum, we hope to see
further studies along this direction in other models and/or contexts as well.
Presumably, similar to the resolution of the `two-band superconductivity 
without supercurrent' controversy near the Dirac points in 
graphene~\cite{kopnin08, kopnin10}, such effects may already be playing 
a part in the multi-band family of high-$T_c$ superconductors that are 
waiting to be uncovered and characterized.

\begin{acknowledgments}
The author acknowledges funding from T{\"U}B{\.I}TAK and the BAGEP award 
of the Turkish Science Academy.
\end{acknowledgments}

\appendix

\section{Usual Hubbard Model}
\label{sec:app}

For the sake of completeness, here we included the self-consistent 
solutions of Eqs.~(\ref{eqn:op})-(\ref{eqn:bkt}) for $\alpha = 0$
or equivalently $\alpha = 1/1$, where
$
\varepsilon_\mathbf{k} = -2t \cos(k_x a) - 2t \cos(k_y a).
$
As we noted in Sec.~\ref{sec:numerics}, even though $\Delta$ 
and $T_{BKT}$ must, in theory, vanish strictly at $U = 0$ wherever 
$\mu$ lies within the band, i.e., $-4t < \mu < 4t$, this appears 
not to be the case in Fig.~3 as well, e.g., $U \to 0$ regions appear white 
instead of black. This is again due to a lack of our numerical resolution 
as the non-linear solver fails to converge once the relative accuracy of two
consecutive $\Delta$ iterations reduces below the order of $10^{-5}$. 
Despite this shortage, we find 
$k_B T_c^{max} \approx 0.2528t$ at $\mu = 0$ when $U \approx 3.75t$,
which is in very good agreement with an earlier estimate~\cite{denteneer93}.

\begin{figure}[htbp]
\includegraphics[scale=0.51]{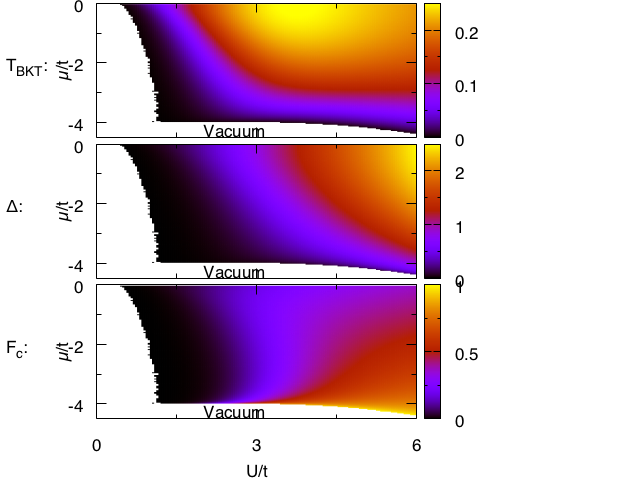}
\label{fig:11}
\caption{(color online)
The critical SF transition temperature $k_B T_{BKT}/t$  is shown in the first row
together with the corresponding SF order parameter $\Delta/t$ in the second row,
and condensate fraction $F_c/F$ in the last row.
}
\end{figure}


\begin{thebibliography}{99}
\bibitem{umucalilar17note}
The extra $q$ factor appearing in the denominator of Eq.~(3) is a typo
in Ref.~\cite{umucalilar17}.
\bibitem{bktnote}
If $D_{\mu \nu}$ is not isotropic then the universal relation becomes
$
k_B T_{BKT} = (\pi/8) \det \mathbb {D}.
$
\bibitem{gapbcs} Even though the mean-field BCS theory breaks down 
when $T > T_{BKT}$, we find that
$
\Delta = \sqrt{3} \Delta_0 \sqrt{1-T/T_{BCS}}
$
works extremely well for $T \approx T_{BCS}$. Here, $\Delta_0$ is the 
$T = 0$ order parameter, and the pre-factor $\sqrt{3}$ is quite robust 
in the entire parameter range. See also~\cite{iskin17} for a similar observation.
\end{thebibliography}
\end{document}